\documentclass[pre,showpacs,twocolumn]{revtex4}
\usepackage{amsmath}
\usepackage{amsfonts}
\usepackage[final]{graphicx}

\def\qdebye{q_{\text{D}}}

\begin{document}

\title{Stiff-Glass Approximation of Mode-Coupling Theory}
\author{Th.~Voigtmann}
\affiliation{Physik-Department, Technische Universit\"at M\"unchen,
             85747 Garching, Germany}

\date{\today}

\begin{abstract} 
The stiff-glass approximation of the mode-coupling theory of the
glass-transition---as introduced in a recent paper by G\"otze and
Mayr for a discussion of phenomena resembling the Boson Peak and
High-Frequency Sound observations made in real glass-formers---is examined
in detail. It amounts to a neglection of two-phonon processes
and thus provides a clear physical picture for the MCT
solutions deep in the glass. In addition, the effect of
additional symplifying approximations and the combination
of these approximations is studied.
\end{abstract}

\maketitle

\section{Introduction}

In a recent paper, G\"otze and Mayr \cite{Goetze2000} demonstrated that
the mode-coupling theory for the glass transition (MCT) applied to a model
of the hard-sphere system (HSS) at relatively high packing fractions
yields spectra for the density correlation functions, that display a broad
asymmetric peak as well as a sound-like mode at rather high wave vectors.
Similar observations have been made in real glass formers and are known as
the Boson Peak (BP) and high-frequency sound (HFS), respectively.
To investigate these phenomena in detail, a large number of experiments has been
performed, using various techniques such as Raman spectroscopy,
x-ray spectroscopy, neutron scattering, or molecular-dynamics simulations.
For an account of the literature, the reader is referred to the papers cited
in \cite{Goetze2000}.
This work also contains a summary of the features signifying both the
HFS and the BP.

The mentioned aspects of the full numerical MCT solutions can be understood
on the
basis of two additional approximations; one being the generalized hydrodynamic
(GHD) approximation, the other being the so-called stiff-glass approximation
(SGA). Performing these two approximations, one arrives at an equation
that, while it still has to be solved numerically, provides physical insight
on the origin of the BP- and HFS-like features of the HSS model.
Especially the SGA can be understood on physical grounds as a description
of phonon scattering from the frozen-in amorphous structure, neglecting
two-phonon processes.
In \cite{Goetze2000} a further simplification was carried out.
It amounts to dropping the remaining wave-vector dependence of the microscopic
frequencies in the GHD-SGA formula. As it leads to an equation that has
been discussed earlier as a ``schematic'' toy model of the MCT, we shall
call it the schematic-model description in the following. One obtains an
analytically solvable expression that, despite its simplicity, gives a
good description of the memory kernel responsible for the BP-like spectra.

The aim of this paper is to drop this last simplification and to consider
the SGA equations directly. This way, all of the additional simplifications
applied on top of MCT can be investigated independently. We shall also
compare with the analytic solution of \cite{Goetze2000} to understand
its range of applicability. The discussion is meant to enlighten the
physical mechanism behind the MCT boson peak, since the SGA is connected
with a clear-cut physical picture.

The paper is organized as follows:
Section~\ref{sec:eom} will present a summary of the basic equations and
approximations outlined above.
In Sec.~\ref{sec:sga}, the solutions of the SGA equations with the
full $q$ dependence taken into account will be presented.
Section~\ref{sec:ghd} examines the effect of the GHD and schematic
approximations, and Sec.~\ref{sec:conclusio} offers some conclusions.

\section{Equations of Motion}\label{sec:eom}

\subsection{Mode-Coupling Theory}

The mode-coupling theory of the glass transition (MCT) describes the dynamics
of a glass-forming liquid in terms of the density correlators
$\Phi_q(t)=\langle\varrho_{\vec q}^*(t)\varrho_{\vec q}\rangle$. Here,
$\varrho_{\vec q}=\sum_j\exp(i\vec q\vec r_j)/\sqrt N$ are the density
fluctuations of wave vector $\vec q$, for a system of $N$ particles at density
$\rho$. $\langle\dots\rangle$ denotes canonical averaging.
For amorphous systems, $\Phi_q(t)$ depends on the wave-vector
only through its modulus $q=|\vec q|$. At short times, one has
$\Phi_q(t)=S_q-\frac12S_q\Omega_q^2t^2+\dots$, where $S_q=\langle|\varrho_
{\vec q}|^2\rangle$ is the static structure factor specifying the equilibrium
structure of the system. $\Omega_q^2=v^2q^2/S_q$ are the phonon frequencies.
A system of hard spheres shall be considered.
The units of time and length are chosen in accordance with
\cite{Goetze2000} such that the thermal velocity
$v=2.5$ and the sphere diameter $d=1$.

The normalized density correlators $\phi_q(t)$, given by $\Phi_q(t)=
S_q\phi_q(t)$, obey the equations of motion obtained via a Mori-Zwanzig
projection operator formalism,
\begin{subequations}\label{eom}
\begin{equation}\label{eomt}
  \ddot\phi_q(t)+\Omega_q^2\phi_q(t)+\Omega_q^2\int_0^t
  m_q(t-t')\dot\phi_q(t')\,dt'=0\,.
\end{equation}
Here $m_q(t)$ is the memory kernel, a correlation function of fluctuating
forces. If one introduces Fourier-Laplace transformed quantities
with the convention $\phi_q(z)=i\int_0^\infty\exp(izt)\phi_q(t)dt$,
Eq.~(\ref{eomt}) is equivalent to
\begin{equation}\label{eomchi}
  \chi_q(z)/\chi_q^0=-\Omega_q^2/\left[z^2-\Omega_q^2+\Omega_q^2zm_q(z)\right]
  \,,
\end{equation}
\end{subequations}
where the fluctuation dissipation theorem has been used to connect
$\phi_q(z)$ to the dynamical susceptibility $\chi_q(z)$:
$\chi_q(z)/\chi_q^0=z\phi_q(z)+1$, with $\chi_q^0$ denoting the
isothermal compressibility.

Within MCT, the memory kernel $m_q(t)$ is written as the sum of a regular
contribution and a so-called mode-coupling contribution, the latter
describing the slow dynamics important close to the glass transition.
For details, the reader is referred to the original publications
\cite{Bengtzelius1984,Leutheusser1984} and standard literature
\cite{Goetze1991b}.
One gets for $m_q(t)$ a quadratic functional in $\phi(t)$,
$m_q^{\text{MCT}}(t)={\mathcal F}_q[\phi]$,
\begin{subequations}\label{mem}
\begin{equation}
  m^{\text{MCT}}_q(t)
  =\int_{\vec p=\vec q-\vec k}\frac{d^3k}{(2\pi)^3}
    V(\vec q\vec k\vec p)\phi_k(t)\phi_p(t)\,,
\end{equation}
with coupling coefficients given by the equilibrium structure of the system,
\begin{equation}
  V(\vec q\vec k\vec p)=\rho S_qS_kS_p[(\vec q\vec k)c_k
  +(\vec q\vec p)c_p]^2/(2q^4)\,.
\end{equation}
Here $c_q$ denotes the Ornstein-Zernike direct correlation function,
connected to the static structure factor by $S_q=1/(1-\varrho c_q)$.
If one introduces a wave-vector grid and approximates the integral as a
Riemann sum, the mode-coupling functional can be written as
${\mathcal F}_q[f]=\sum_{kp}V_{qkp}f_kf_p$, where $k$ and $p$ run over the
set of all grid points. The coefficients $V_{qkp}$ are trivially related
to the $V(\vec q\vec k\vec p)$.
\end{subequations}

For the discussion of the MCT dynamics in the boson-peak region,
G\"otze and Mayr \cite{Goetze2000} chose a model of the hard-sphere system,
given by evaluating $S_q$ in the Percus-Yevick approximation, introducing
a grid of $M=300$ wave vectors $q$ with a cutoff of $q^*=40$, and by
neglecting the regular part of the memory kernel.
The dependence of the solutions on these approximations will have to be
investigated seperately.
For the structure factor used, a cutoff dependence of the solutions cannot
be neglected, as was already mentioned in \cite{Goetze2000}, and one is lead
to the conclusion that the model described here is one of a soft-core
fluid rather than true hard spheres. Even more so, the influence of the
regular term is unclear at present. Recent calculations indicate that,
at least in the white-noise approximation based on a generalized Enskog theory,
the regular damping might be so large as to render the BP part of the
spectrum nonexistent for a true hard-sphere system \cite{Sperl.priv}.
The question, however, remains unanswered still for regular potentials.
Nevertheless, since the concern of this paper is a comparison
with the results presented in \cite{Goetze2000}, we shall choose all
numerical parameters the same. Equations (\ref{eom}) and (\ref{mem}) are
closed then, and since $S_q$ for the hard-sphere system does not depend on
temperature, the only control parameter is the density $\rho$, which
shall be written as the packing fraction $\varphi=\pi\rho/6$.

\subsection{Stiff-Glass Approximation}

The starting point for a theoretical understanding of the MCT high-frequency
dynamics is a reformulation of the equations of motion,
Eqs.~(\ref{eom}), within the glass state.
One introduces new correlators $\hat\phi_q(t)$ that only
deal with the decay relative to the frozen structure given by the
nonergodicity parameter $f_q$,
 $ \phi_q(t)=f_q+(1-f_q)\hat\phi_q(t) $.
Equations (\ref{eom}) are covariant in the sense that the same equations
hold for the new variables,
\begin{equation}\label{hatchi}
  \chi_q(z)/\hat\chi^0_q=-\hat\Omega_q^2/\left[z^2-\hat\Omega_q^2
  (1-z\hat m(z))\right]\,,
\end{equation}
provided one replaces the static susceptibility by
$\hat\chi^0_q=\chi^0_q\,(1-f_q)$,
the frequencies $\Omega_q^2$ by
\begin{equation}
  \hat\Omega_q^2=\Omega_q^2/(1-f_q)\,,
\end{equation}
and the memory kernel by
\begin{subequations}
\begin{align}
  \hat m_q(t)&=\hat{\mathcal F}_q^{(1)}[\hat\phi]
              =\hat{\mathcal F}_q^{(1)}[\hat\phi]
              +\hat{\mathcal F}_q^{(2)}[\hat\phi]\,,\\
  \hat{\mathcal F}_q^{(1)}[\hat f]&=\sum_k\hat V_{qk}\hat f_k\,,\\
  \hat{\mathcal F}_q^{(2)}[\hat f]&=\sum_{kp}\hat V_{qkp}\hat f_k\hat f_p\,,
\end{align}
\end{subequations}
with new coupling coefficients given by
\begin{subequations}
\begin{align}
  \hat V_{qk}&=2(1-f_q)\sum_p V_{qkp}f_p(1-f_k)\,,\\
  \hat V_{qkp}&=(1-f_q)V_{qkp}(1-f_k)(1-f_p)\,.
\end{align}
\end{subequations}

The new variables $\hat\phi_q(t)$ and $\hat m_q(t)$ have the properties
that their long-time limits vanish, thus the low-frequency limits of their
Laplace transforms are regular.

To study BP phenomena, the limit of high packing fraction is of
interest. There the original coupling coefficients become large, formally
written $V_{qkp}={\mathcal O}(1/\eta)$ with some small parameter $\eta$.
The virtue of the covariant transformation now is that it implies
$1-f_q={\mathcal O}(\eta)$, and thus
\begin{align}
  \hat V_{qk}&={\mathcal O}(\eta)\,,&
  \hat V_{qkp}&={\mathcal O}(\eta^2)\,,
\end{align}
i.e.\ the new coupling coefficients become small.
This suggests to drop the second-order contribution
to the memory kernel $\hat m_q(t)$, such defining the stiff-glass approximation
(SGA), cf.\ Eq.~(14) of \cite{Goetze2000}:
\begin{subequations}\label{sga}
\begin{align}
  \hat\phi^{(1)}_q(z)&=-1/\left[z-\hat\Omega_q^2/\left[z+\hat\Omega_q^2
  \hat m^{(1)}_q(z)\right]\right]\,,\\
  \hat m^{(1)}_q(z)&=\sum_k\hat V_{qk}\hat\phi^{(1)}_k(z)\,.
\end{align}
\end{subequations}

Equations similar to Eq.~(\ref{sga}) were derived in a different context
by other authors.
Let us, for example, rewrite these equations in a notation closer to the
conventions of field theory. The dynamical susceptibility can be rewritten
for positive frequencies $z=\omega^2$, $\omega>0$, into a propagator
$G(q,\omega)=-\hat\chi^{(1)}_q(\sqrt{z})/(\hat\chi_q\hat\Omega_q^2)$,
depending on the bare dispersion relation $\epsilon(q)=\hat\Omega_q^2
(1+\sum_k\hat V_{qk})$, and on the complex self energy $\Sigma(p,\omega)$,
to give
\begin{subequations}\label{dyson}
\begin{align}
  G(q,\omega)&=\frac{1}{\omega-\epsilon(q)-\Sigma(q,\omega)}\,\\
  \Sigma(q,\omega)&=\int dk \tilde V_{qk} G(k,\omega)\,,
\end{align}
\end{subequations}
where $\tilde V_{qk}=\hat\Omega_q^2\hat V_{qk}\hat\Omega_k^2$ are
positive coupling constants. This
is equivalent to the equations discussed by Mart{\'\i}n-Mayor \emph{et~al.}
\cite{MartinMayor2001}
and Grigera \emph{et~al.}
\cite{Grigera2001} in connection with Euclidean random matrix theory, cf.\
Eqs.~(13) and (17) of the latter paper, where, however, the vertices
$\tilde V_{qk}$ are different.
Equations (\ref{dyson}) can be viewed as a special case of the Dyson equation
\cite{Economou1983}.

Equations (\ref{sga}) still have to be solved numerically.
Thus, G\"otze and Mayr
\cite{Goetze2000} proposed two additional simplifications, the first of
which is the Generalized Hydrodynamic (GHD) description. One replaces the
memory kernel $\hat m^{(1)}_q(z)$ by its $(q\to0)$ limit,
$K(z)=\hat m^{(1)}_{q=0}(z)$. If one is only interested in a description
of sound modes at small $q$, one could also replace the frequencies
by their leading-order contributions, $\hat\Omega_q=v_\infty q
+{\mathcal O}(q^2)$, with the high-frequency sound velocity
$v_\infty=v/\sqrt{(1-f_0)S_0}$. In the following, we will only discuss
the GHD-SGA obtained by keeping the full $q$-dependence of $\hat\Omega_q$,
\begin{equation}\label{sgaghd}
  \hat\chi^{\text{GHD}}_q(z)/\hat\chi_q=-\hat\Omega_q^2/
  \left[z^2-\hat\Omega_q^2(1-zK(z))\right]\,.
\end{equation}

A further simplification was obtained in \cite{Goetze2000} by replacing
$\hat\Omega_q$ with some $q$-independent averaged value $\tilde\Omega$. This
was motivated by the observation that in the range around the first diffraction
peak in the structure factor, a $q$ range known to constitute the dominant
contribution to the mode-coupling kernel \cite{Bengtzelius1984},
the variation of the renormalized values $\hat\Omega_q$ with $q$ is
suppressed relative to that of $\Omega_q$.
As a result, one arrives at
\begin{subequations}\label{schematic}
\begin{align}
  \label{phi_f1}
  \tilde\phi(z)&=-1/\left[z-\tilde\Omega^2/\left[z+\tilde\Omega^2
    \tilde K(z)\right]
  \right]\,,\\
  \label{K_f1}
  \tilde K(z)&=w_1\tilde\phi(z)\,,
\end{align}
with $w_1=\int dk\,\hat V_{0k}$.
The set of Eqs.~(\ref{phi_f1},\ref{K_f1}) constitutes a well-known schematic
model of MCT discussed earlier
\cite{Goetze1984b,Jacobs1986,Goetze1991b}, the so-called ${\mathrm F}_1$ model.
\end{subequations}
It can be solved trivially for either $\tilde\phi(z)$ or $\tilde K(z)$.
With the notation $z=\omega+i0$ we get
\begin{equation}\label{semi-ellipse}
  \tilde K(\omega)\approx\left[\omega_+\omega_--\tilde\omega^2+
  \sqrt{\tilde\omega^2-\omega_-^2}\sqrt{\tilde\omega^2-\omega_+^2}\right]
  /(2\omega)\,,
\end{equation}
with $\omega_\pm=1\pm\sqrt{w_1}$, and $\tilde\omega^2=\omega^2/\tilde\Omega^2$.
The ``semi-ellipse''
spectrum $\tilde K''(\omega)$ thus obtained was demonstrated in
\cite{Goetze2000} to have striking similarities to the spectra of the
full MCT solutions in the BP regime at $\varphi=0.6$.

Eq.~(\ref{semi-ellipse}) implies a ``gap'' in the kernel spectrum for
$0<\omega<\omega_-$. The full solutions show a pseudo-gap for frequencies
below the BP frequency resembling this finding. The smearing out of the gap
due to two-phonon processes can be understood on the basis of the schematic
model by including these processes in a perturbative manner,
which amounts to replacing $\tilde K(\omega)$ in Eq.~(\ref{phi_f1}) with
$\tilde K^{(2)}(\omega)=\tilde K(\omega)+K^{(2)}(\omega)$,
where the latter term reads \cite{Goetze2000}
\begin{equation}\label{k2}
  K^{(2)}(\omega)
  =\frac1\pi\frac{w_2}{w_1^2}\int\tilde K(\omega-\omega')
   \tilde K''(\omega')\,d\omega'\,,
\end{equation}
with the integrated second-order coupling coefficient
$w_2=\int dk\,\hat V_{0kk}$.
$\tilde K^{(2)}(\omega)$ is then still given by Eq.~(\ref{semi-ellipse}),
but one has to set $\tilde\omega^2=\omega^2/\tilde\Omega^2+K^{(2)}(\omega)$.
Analytic solvability can be regained by evaluating this in the
$(\omega\to0)$ limit, $K^{(2)}(\omega\to0)=i\tau$,
\begin{equation}\label{k2tau}
  \tau=\frac{w_2}{\sqrt{w_1}}\frac{8}{3\pi\tilde\Omega}\,.
\end{equation}

Since arriving at Eq.~(\ref{semi-ellipse}) involved three subsequent
approximations which, in principle, could be applied independently from
one another, it seems worthwile to discuss the effects of each approximation
by itself. Especially going over to the schematic model description of the
memory kernel as a last step is, in contrast to the GHD-SGA, uncontrolled in
the sense that its derivation is not based on a well-defined mathematical
procedure.

The GHD approximation was shown in \cite{Goetze2000}
to be of very good quality, provided on does apply it to the transformed
memory kernel $\hat m_q(z)$. In other words, keeping the full $q$-dependence
of the nonergodicity parameters $f_q$ is of vital importance for a description
of the BP and HFS spectra. It can thus be expected to be of similar
quality when applied together with the SGA.

While not leading to analytically solvable expressions, the SGA
alone, Eqs.~(\ref{sga}), has a clear-cut physical meaning:
The first-order contribution to the memory kernel can be interpreted as
an elastic scattering of phonons due to the frozen disorder in the glass,
a mechanism thought to be dominant over two-phonon decay processes, which
are represented by the second-order term and therefore ignored in the
SGA. It should be noted, however, that this does not correspond to a
harmonic approximation.

\section{Solution of the SGA equations}\label{sec:sga}

To obtain numerical solutions of Eqs.~(\ref{sga}), an algorithm adapted from
the one used in \cite{Goetze2000} was employed. It solves the initial-value
problem for the coupled integro-differential equations corresponding to
Eqs.~(\ref{sga}) in the time domain. Regarding the stability of this
method, proofs are available \cite{Goetze1995b},
which apply to the SGA as a special case. Here it merely is important
that the mode-coupling vertices $\hat V_{qk}$ are positive.
The solutions were Fourier transformed
to give the spectra $\phi_q''(\omega)$ and $m_q''(\omega)$.

\begin{figure}
\includegraphics[width=.9\linewidth]{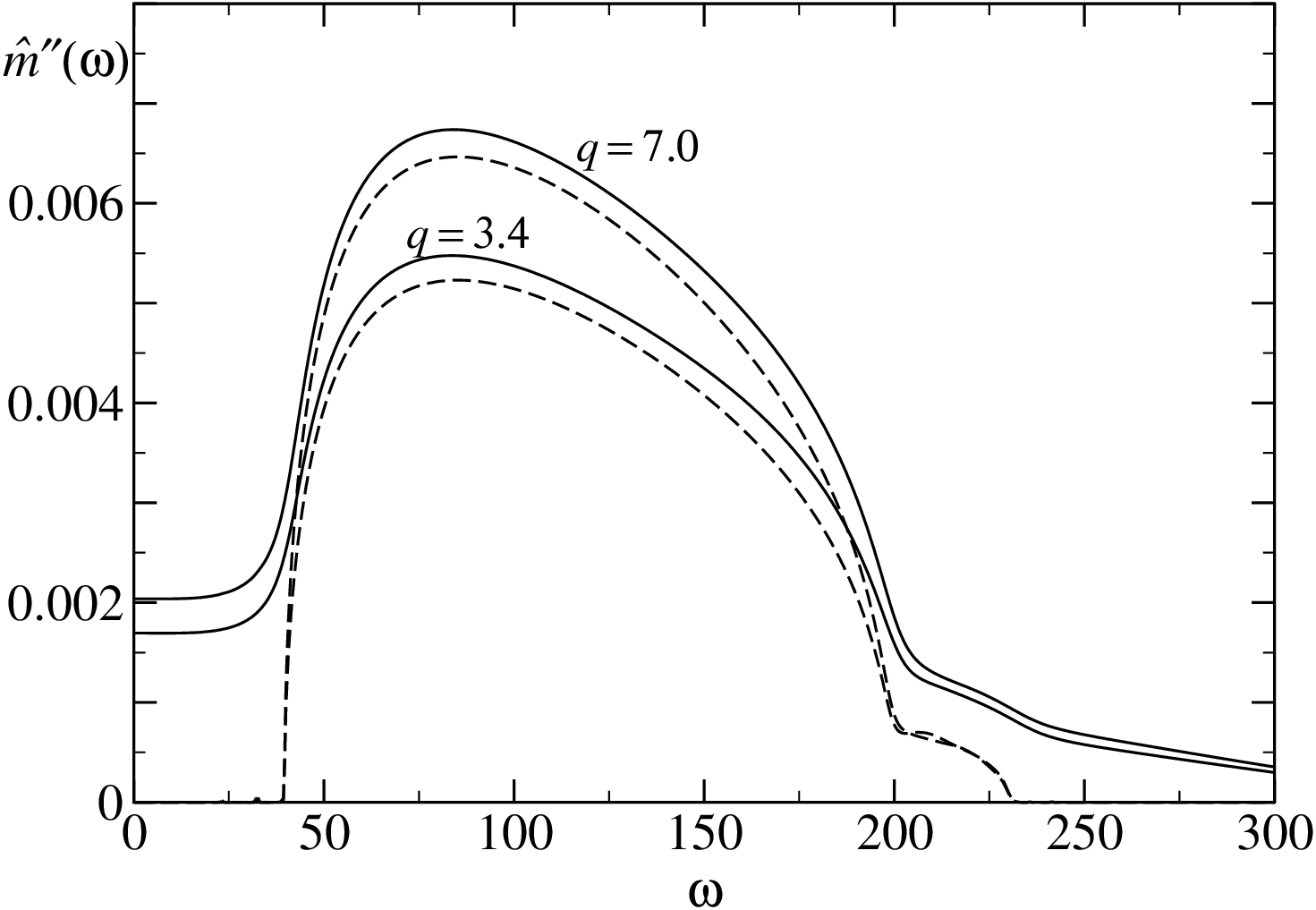}
\caption{\label{fig1}Transformed memory-kernel spectra $\hat m''_q(\omega)$
of the full-MCT solutions for the HSS model, Eq.~(\protect\ref{hatchi}), for
wave vectors $q=3.4$, and $q=7.0$,
at packing fraction $\varphi=0.6$ (solid lines).
The dashed lines show the corresponding spectra obtained
within the stiff-glass approximation (SGA), Eq.~(\protect\ref{sga}).}
\end{figure}

Let us start the discussion by a comparison of the memory kernel spectra
$\hat m_q''(\omega)$, obtained from the solution of the full MCT equations,
Eq.~(\ref{hatchi}), and of the SGA equations, Eq.~(\ref{sga}). Figure~\ref{fig1}
shows such a comparison for packing fraction $\varphi=0.6$.
One notes a dominant broad and asymmetric peak responsible for the
BP features of the correlator spectra \cite{Goetze2000}. The SGA solutions
still recover the shape of that peak reasonably well; an observation together
with its form being reminiscent of the semi-ellipse known from the schematic
model further justifying the analogy put forward by the schematic model
description.
At frequencies $\omega\lesssim40$, a pseudo-gap can be seen. As discussed in
\cite{Goetze2000}, the SGA spectrum is not exactly zero within this gap;
rather one obtains Rayleigh's law,
$\hat m^{(1)\prime\prime}_q(\omega)=R_q\omega^2+{\mathcal O}(\omega^4)$.
The prefactor
of the Rayleigh contribution is, however, so small that it is
invisible in the plot shown. Comparing with the full MCT solution,
one clearly notes the filling of the gap due to the two-phonon modes.

On the high-frequency wing, the spectra obtained from the full solutions
show a nontrivial decay to zero, including, for larger $q$ vectors, a second
small peak at frequencies $200\lesssim\omega\lesssim250$. Interestingly
enough, this peak is also reproduced by the SGA memory kernels.

\begin{figure}
\includegraphics[width=.9\linewidth]{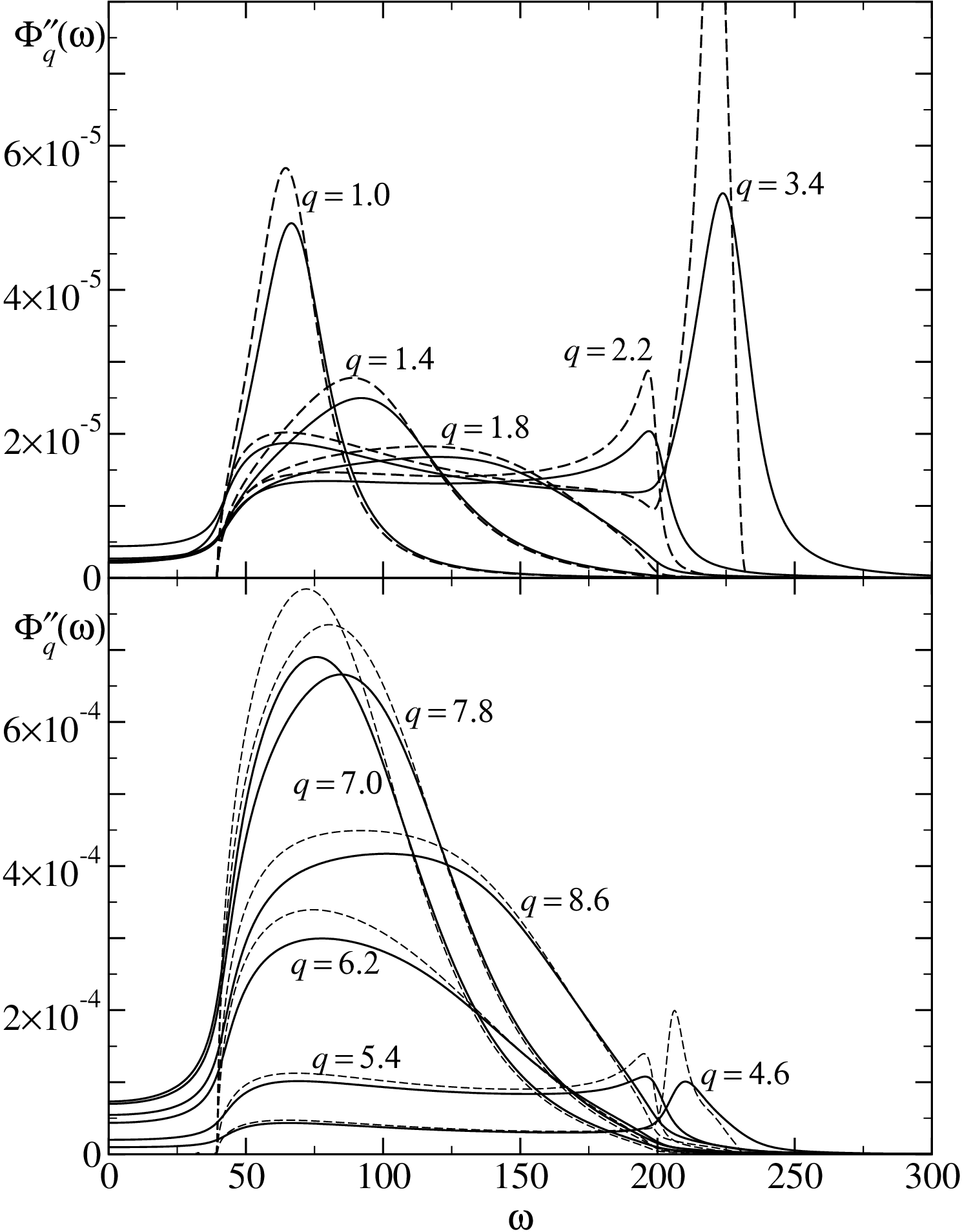}
\caption{\label{fig2}Dynamical structure factor $S(q,\omega)=S_q
\phi''_q(\omega)$ for different wave vector $q$ as labeled at packing fraction
$\varphi=0.6$. The solid
lines show the solutions obtained from the full equations, Eqs.~(\ref{eom}).
The dashed lines are the corresponding SGA spectra, Eqs.~(\ref{sga}).}
\end{figure}

Since the SGA reproduces the memory kernel of the full MCT solutions resonably
well in the frequency range of importance for the discussion of
BP-like phenomena, it comes as no surprise that it also gives a basically
correct description of the density correlator spectra $\Phi_q''(\omega)$.
Figure~\ref{fig2} compares these spectra to those of the full MCT equations.
The BP spectra are, with some quantiative differences, largely correct within
the SGA, and the same holds for the HFS modes for those $q$ vectors, where
the sound damping is dominated by the BP background. At $q\sim\qdebye$,
where $\qdebye\approx4$ is the Debye frequency \cite{Goetze2000}, the
SGA sound-mode damping clearly is too small. This failure is
to be expected from the
above discussion of the memory kernel, since in this case the sound resonance
maximum occurs at frequencies above the BP spectrum, where the SGA memory
kernel spectrum is too small. Similarly, the pseudo-gap at low
frequencies and the rapid decay at large $\omega$ within the SGA are a result
of the same features of the memory kernel spectra.

\begin{figure}
\includegraphics[width=.9\linewidth]{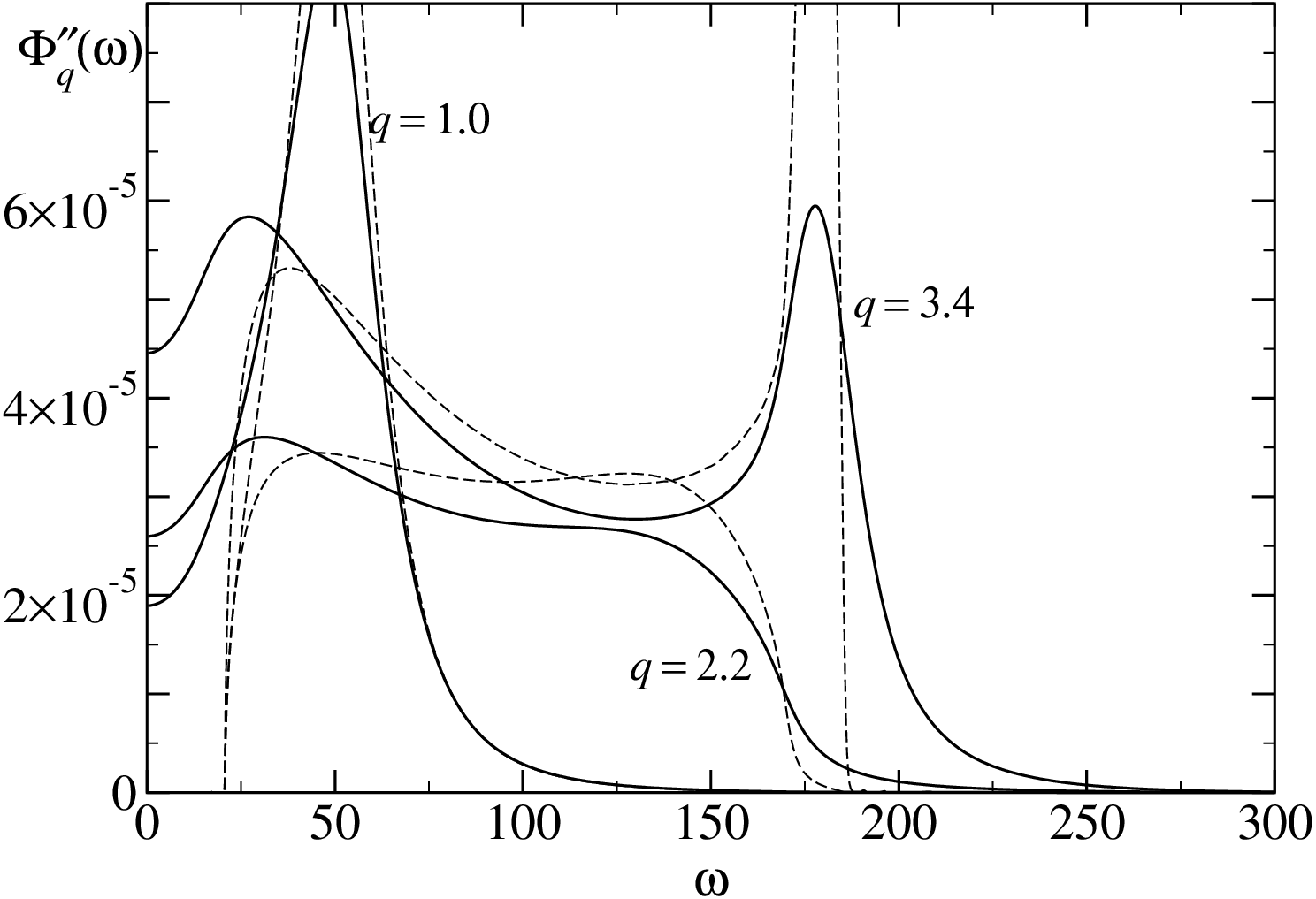}
\caption{\label{fig3}Comparison of full MCT and SGA for spectra
$\Phi''_q(\omega)$ as in Fig.~\protect\ref{fig2}, but for $\varphi=0.5676$,
and wave vectors $q=1.0$, $2.2$, and $3.4$. Other $q$ were left out to avoid
overcrowding the figure.}
\end{figure}

\begin{figure}
\includegraphics[width=.9\linewidth]{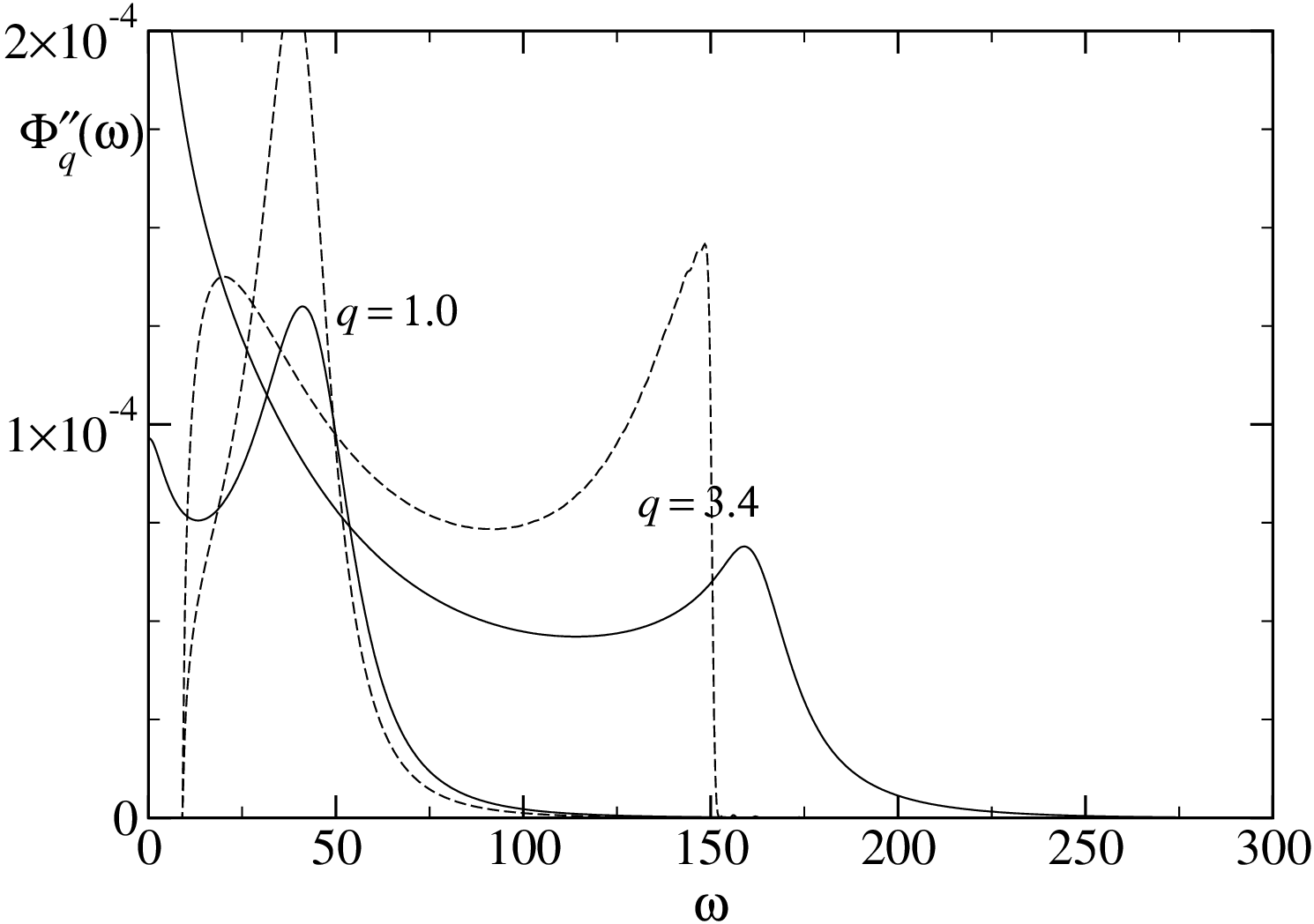}
\caption{\label{fig4}Comparison of full MCT and SGA for spectra
$\Phi''_q(\omega)$ as in Fig.~\protect\ref{fig2}, but for $\varphi=0.54$,
and wave vectors $q=1.0$ and $q=3.4$.}
\end{figure}

The SGA is based on the two-mode contributions of order
${\mathcal O}(\eta^2)$ being neglegible in comparison to one-mode contributions
of order ${\mathcal O}(\eta)$. The quality of this approximation
is expected to decrease with lowering the packing fraction, since $\eta$
increases with decreasing $\varphi$.
Figures~\ref{fig3} and \ref{fig4} repeat the
comparison made in Fig.~\ref{fig2}, now for packing fractions $\varphi=0.5676$
and $\varphi=0.54$, respectively. One can identify the increasing intensity
for low frequencies, eventually becoming so strong as to dominate over the
BP background for $\varphi=0.54$.
Nevertheless, the SGA gives the qualitatively correct picture even for
$\varphi=0.54$ at frequencies $\omega\gtrsim25$, the number depending somewhat
on the wave vector.
One also notices the shift of both the sound resonance and the BP maximum
to lower frequencies as the packing fraction decreases. This further
demonstrates two features typical for the BP and the HFS \cite{Goetze2000}.

\begin{figure}
\includegraphics[width=.9\linewidth]{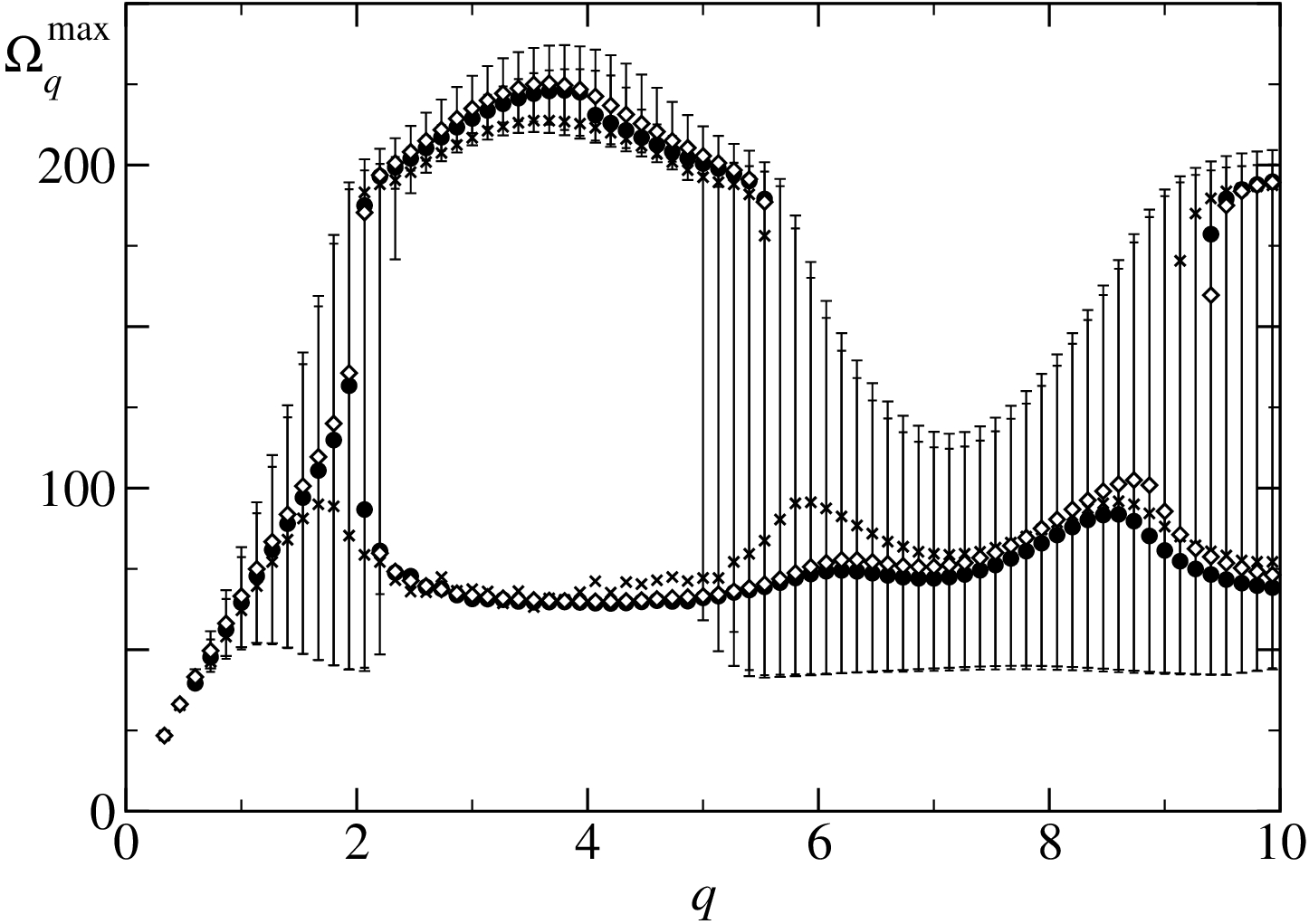}
\caption{\label{fig5}Maximum positions of the density-correlator spectra
$\phi''_q(\omega)$ obtained from the full MCT solutions at $\varphi=0.6$
versus wave vector $q$ (open diamonds). Two symbols are drawn where two distinct
maxima (separated by a minimum) could be identified. Filled circles indicate the
maximum positions obtained within the SGA, Eq.~(\protect\ref{sga}), while
crosses correspond to the results of the SGA-GHD, Eq.~(\protect\ref{sgaghd}).
The bars indicate the full widths at half maximum (FWHM) as described
in the text. The smaller bars are FWHM of the SGA solutions, FWHM data
for the SGA-GHD is left out to enhance readability.}
\end{figure}

\begin{figure}
\includegraphics[width=.9\linewidth]{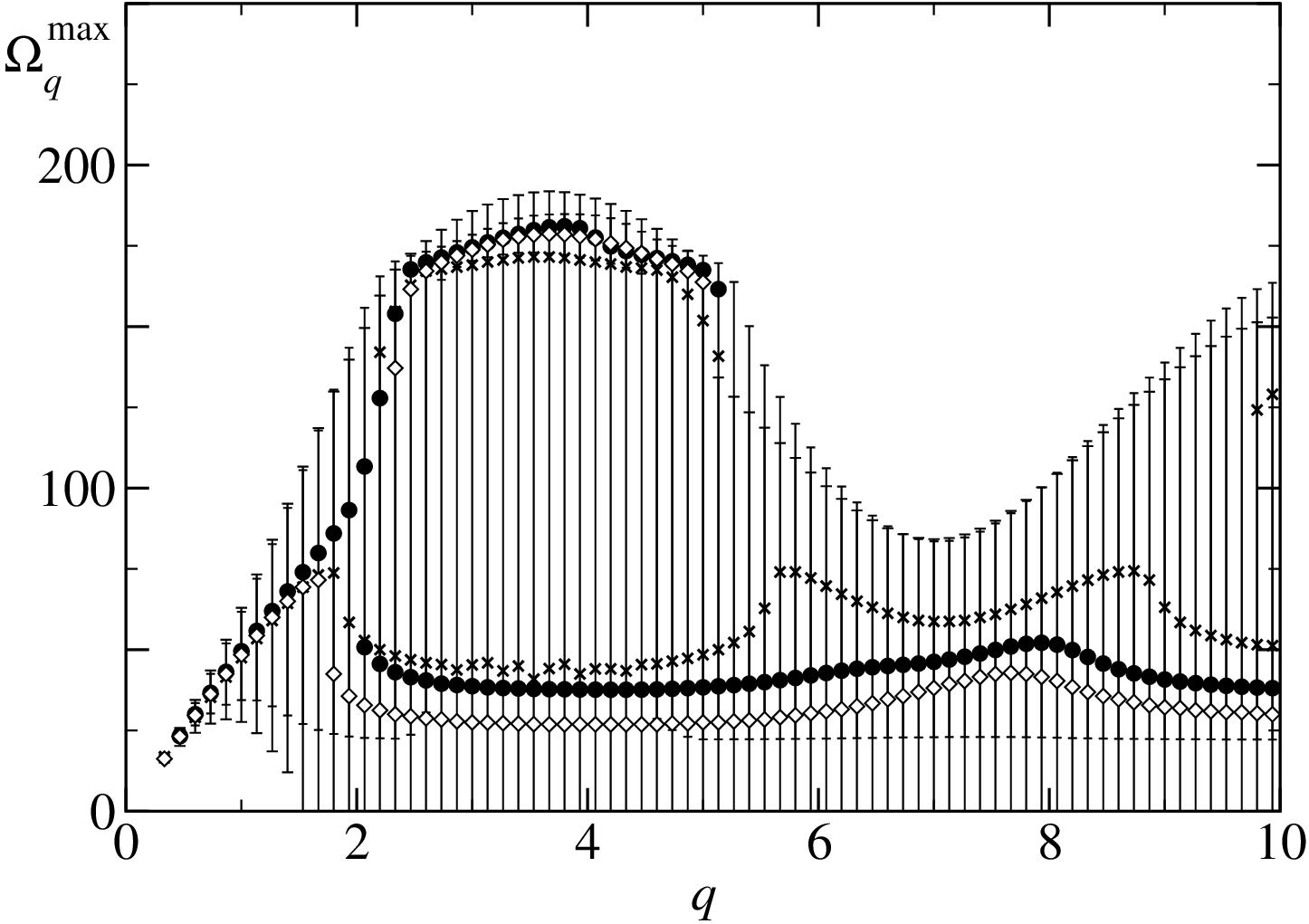}
\caption{\label{fig6}Same as Fig.~\protect\ref{fig5}, but for packing fraction
$\varphi=0.5676$.}
\end{figure}

One notes that for $\varphi=0.54$ and $\varphi=0.5676$ the true HFS maximum
positions are higher than those read off the SGA solutions, especially for
$q$ vectors where the SGA significantly underestimates the damping. This
effect is also known as level repulsion. For a discussion of the effects of
the SGA on the HFS modes, Figs.~\ref{fig5} and \ref{fig6}
show the maximum positions of the spectra $\phi_q''(\omega)$ and
the resonance widths versus wave vector $q$ for different packing fractions.
Two maximum positions are noted whenever the BP background and the HFS mode
could be identified as distinct, separated by a spectral minimum.
The full widths at half maximum (FWHM) were obtained
by marking the lowest and the highest frequency where the correlator spectrum
reaches half the value of its global maximum.

For the HFS maxima as well as the BP maxima at $\varphi=0.6$, Fig.~\ref{fig5},
good agreement between the full solutions, reproduced from \cite{Goetze2000},
and the SGA solutions is found. This even holds in the low-$q$ regime, where
only a damped sound mode can be seen in the spectra. Here the shift in the
maximum position due to damping is still small, such that an underestimation of
the damping by the SGA does not necessarily lead to a wrong description of the
position. The agreement of the FWHM values also is good, with the reservation
that the SGA values are too small in general. The lower frequency for which
the spectrum reaches half the maximum intensity is described better than
the higher one within the SGA, since here the pseudo-gap marks the dominant
rise in the spectrum.

Comparing the maximum positions and FWHM at $\varphi=0.5676$, one notes that
for the BP maximum, the agreement between full MCT and SGA still is very
good. Expectedly, the description of the FWHM is worse, mostly due to the
appearance of the pseudo-gap and due to the underestimation of the sound-mode
damping around $\qdebye$.

\begin{figure}
\includegraphics[width=.9\linewidth]{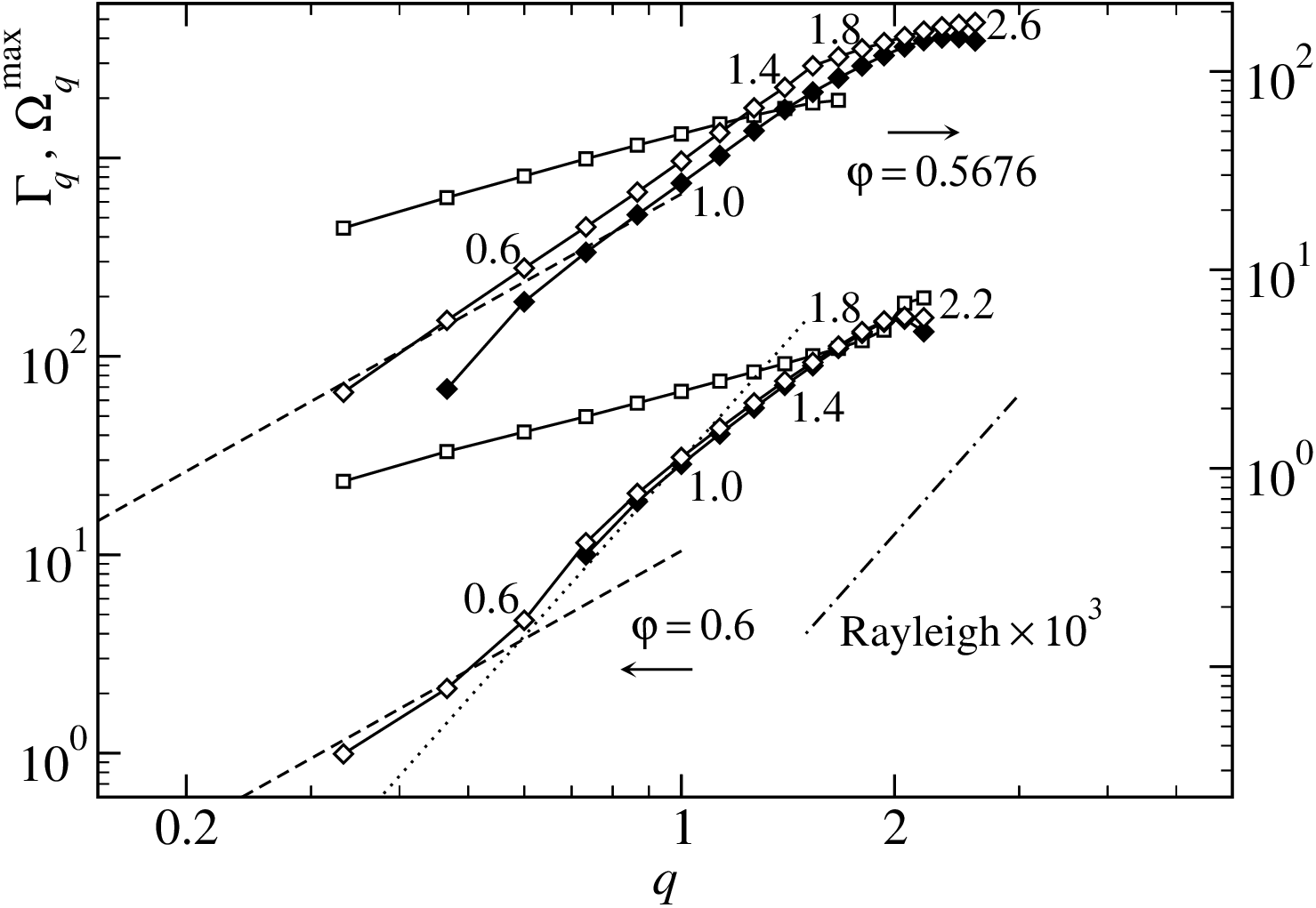}
\caption{\label{fig6add}Double-logarithmic plot of linewidths $\Gamma_q$
(diamonds) and
maximum positions $\Omega_q^{\text{max}}$ (squares)
of the density-correlator spectra
versus wave vector $q$ in the high-frequency-sound regime.
The lower set of data corresponds to packing fraction $\varphi=0.6$ (labels
on left axis). The upper set shows the same data for $\varphi=0.5676$ (labels
on right axis). Open symbols are values determined from the solutions of the
full MCT, filled symbols demonstrate the SGA solutions. The dashed lines
correspond to the low-$q$ asymptote, $\Gamma_q=q^2\gamma$ (see text). The
dotted line corresponds to a $q^4$ power law, while the dot-dashed line
demonstrates the Rayleigh contribution for $\varphi=0.6$, multiplied by a
factor of $10^3$.}
\end{figure}

The sound damping in the regime of the HFS
is illuminated in more detail in Fig.~\ref{fig6add}. Here,
full width at half maximum values $\Gamma_q$ obtained from the full solutions
of Eq.~(\ref{hatchi}) and from within the SGA, Eq.~(\ref{sga}), are compared
for the two packing fractions $\varphi=0.6$ and $\varphi=0.5676$. For
$q\le0.6$ ($0.467$) at $\varphi=0.6$ ($0.5676$), the SGA line widths could
not be determined reliably. This reflects a limitation of the numerical
procedure used, which is not adequate for very weakly damped oscillations,
as occuring within the SGA.
Nevertheless, the errors introduced by the numerical algorithm for low $q$
do not significantly affect the accuracy of the solutions for larger $q$,
since the small-$q$ contribution to the memory kernel is neglegible
\cite{Bengtzelius1984}.

The dashed lines in Fig.~\ref{fig6add} show the low-$q$ asymptote
for the full MCT solutions, $\Gamma_q=q^2\gamma$,
with $\gamma=v_\infty^2\hat m_{q=0}''(\omega\to0)$.
One notices an enhancement above the asymptotic $q^2$ law for $q\gtrsim0.6$.
At around $q=1.4$, the resonance width becomes of the same order of magnitude
as its position, $\Omega_q^{\text{max}}$, and thus a Ioffe-Regel limit is
reached. Above this point, the line shapes differ significantly from a
Lorentzian and hybridization with the BP modes becomes important.
For $\varphi=0.5676$, the maximum positions $\Omega_q^{\text{max}}$ are smaller
than for $\varphi=0.6$, while the damping $\Gamma_q$ changes only weakly
in the regime of the enhancement over the asymptotic law, a behavior also
found in experiment \cite{Masciovecchio1997,Monaco1998}. Only for very small
$q$ vectors, $q\lesssim0.6$, one notices in the full MCT solutions the
expected behavior of $\Gamma_q$.

In the $q$ range from $q\approx0.6$ to $q\approx1.4$, the data for
$\varphi=0.6$ can be approximately described by some power law, with its
exponent being close to $4$. To demonstrate this, the dotted line was included
in Fig.~\ref{fig6add}, representing a $q^4$ law.
While in some investigations, no indication of such a $q^4$ law was
found \cite{Ruocco2000}, it has been argued that this is behaviour is
connected to Rayleigh's law \cite{Rat1999}.
In the present case, however, the approximate $q^4$ power law is
not an indication for Rayleigh scattering. As already noted
in \cite{Goetze2000}, the Rayleigh contribution $R_q$ is much lower. It is
included in
Fig.~\ref{fig6add} as the dot-dashed line. Note that a magnification
by a factor of $1000$ was necessary to make this contribution visible
in the plot. Thus one concludes, that the $q^4$ law for the high-frequency
sound is a mere numerical coincidence.
The possible appearance of a pseudo-power law
for the HFS line width can be seen even more clearly for $\varphi=0.5676$,
where an exponent between $2$ and $4$ would be obtained.

Fig.~\ref{fig6add} again demonstrates the validity of the SGA. At and below
$q\sim0.6$, deviations are to be expected, since the SGA line widths do not
obey the asymptotic $q^2$ law. Rather one expects the SGA damping for $q\to0$
to be given by Rayleigh's law noted above. Despite this fact, the overall
description of $\Gamma_q$ by the SGA is quite good.

\section{Effect of additional approximations}\label{sec:ghd}

\begin{figure}
\includegraphics[width=.9\linewidth]{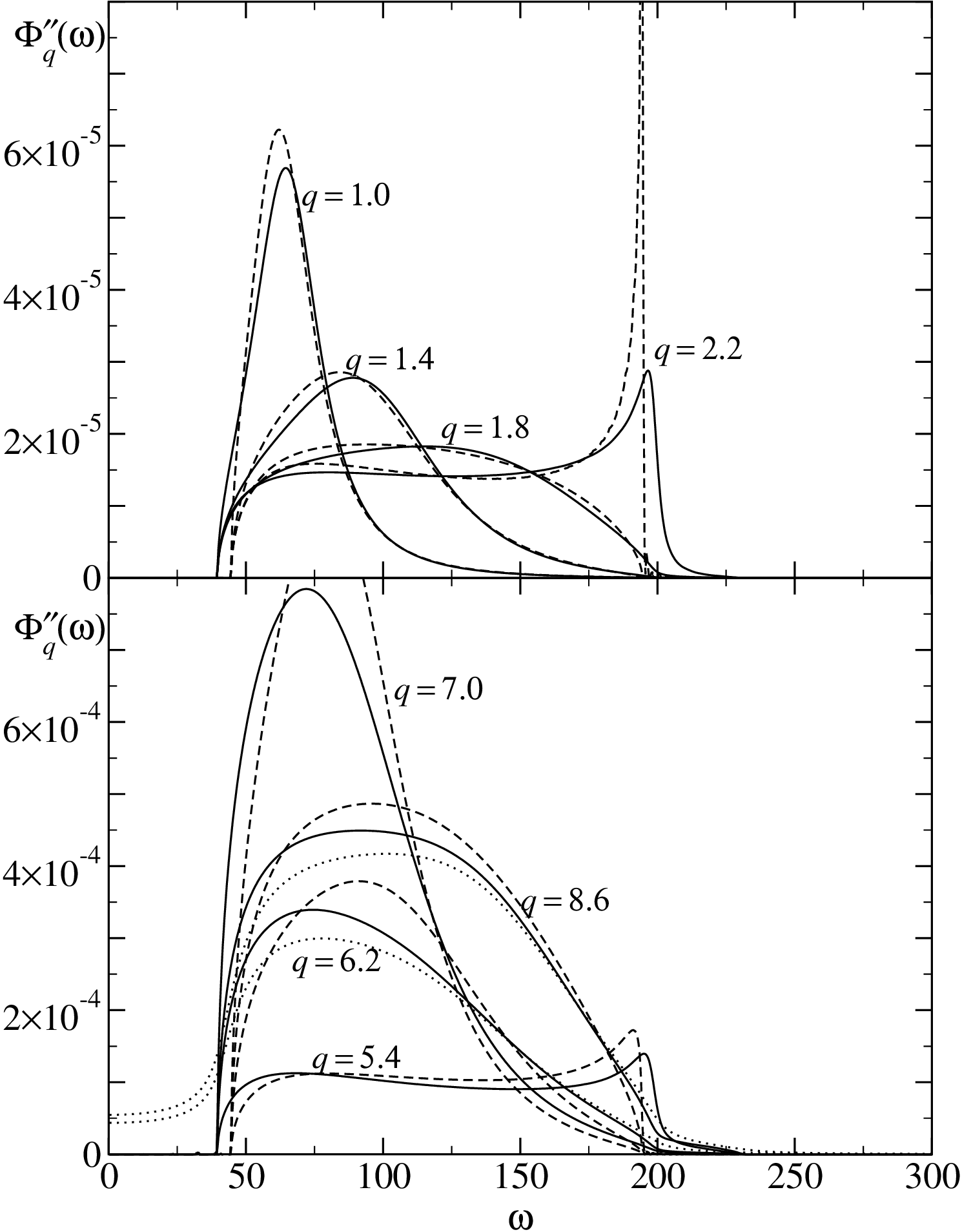}
\caption{\label{fig6a}Spectra for $\varphi=0.6$ and $q$ vectors as labeled
obtained within the SGA (solid lines, reproduced from Fig.~\protect\ref{fig2}).
Dashed lines show the corresponding spectra obtained within the
GHD description in addition to the SGA. The dotted lines show examples
for the full solutions (taken from Fig.~\protect\ref{fig2}).}
\end{figure}

Having discussed the quality of the SGA in comparison to a full MCT solution,
we shall now turn to a discussion of the additional simplifications outlined
in the introduction. Since the $q$ dependence of the memory kernel spectra
is weak, cf.\ Fig.~\ref{fig1}, one is lead to the combination of both the
GHD approximation and the SGA. Figure~\ref{fig6a} shows a comparison of thus
obtained density-correlator spectra with selected full-MCT and SGA spectra
reproduced from Fig.~\ref{fig2}. Indeed, good agreement is found for the
experimentally relevant $q$-vector range at $q\lesssim\qdebye$.

The GHD-SGA solution fails to produce reasonable sound modes for wave vectors
where the position of the sound maximum is at about the upper cutoff for the
BP memory kernel. This is due to the sensitivity of the shape of the memory
kernel in this frequency region. As noted in connection with Fig.~\ref{fig1},
the SGA still reproduces the nontrivial second peak in $m_q''(\omega)$ also
seen in the full MCT solutions. Since this peak only evolves for higher
$q$ vectors, the GHD approximation obviously misses this feature, resulting
in only weakly damped sound modes there. Still, the GHD-SGA combination is
valuable for a discussion of BP-like spectra and the hybridization effects
occuring between HFS modes and the BP spectra. Note that the maximum positions
obtained from the GHD-SGA description show a qualitatively correct behavior.
This can be seen from Figs.~\ref{fig5} and \ref{fig6}.
Within the hybridization regime, the FWHM values (not shown in the figures)
also are almost quantitatively correct.

A comparison between the full solutions and those obtained by applying the
GHD only, still including two-phonon modes, was given in Fig.~6 of
\cite{Goetze2000}. There it can be seen that the GHD approximation is, for
$q\lesssim\qdebye$, at least
as good as the SGA, at the expense of still requiring the computation of
a nonlinear memory kernel.
Only the combination of both SGA and GHD seems to be too
crude to deal with high-frequency sound modes for all wave vectors.
For higher $q$ vectors, the GHD-only approximation of \cite{Goetze2000}
and consequently also the GHD-SGA description somewhat miss the shape of the BP
background spectra, as is evident from the lower panel of Fig.~\ref{fig6a}.
On the other hand, the SGA discussed here regains better quality for those
$q$-vector regions where the HFS mode again shows hybridization with the
BP-like spectrum. This shows that in this region, one-mode contributions from
higher $q$ vectors are important to fully explain the shape of the BP phenomena.

\begin{figure}
\includegraphics[width=.9\linewidth]{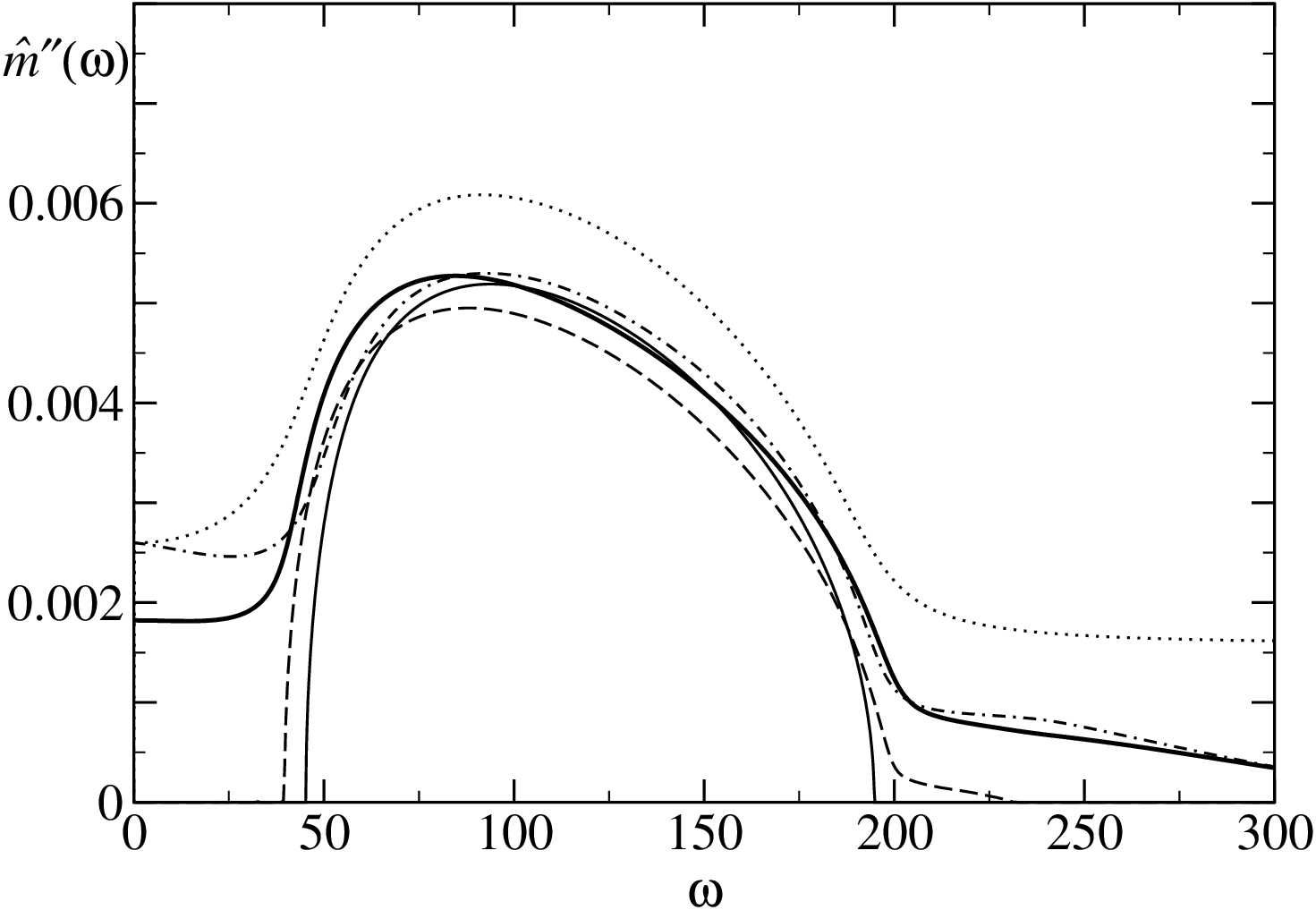}
\caption{\label{fig7}Memory kernel spectrum for $\varphi=0.6$ and $q=0$. The
spectrum $\hat m_{q=0}''(\omega)$
obtained from the full MCT equations, Eq.~(\protect\ref{hatchi}),
is shown as the thick solid line. The dashed line shows the SGA result.
The schematic-model descriptions shown in Fig.~13 of \protect\cite{Goetze2000}
are also reproduced: the thin solid line is the semi-ellipse obtained from
Eq.~(\protect\ref{semi-ellipse}), the dotted line is the spectrum with $\tau$
evaluated from Eq.~(\protect\ref{k2tau}), see text for details.
The dot-dashed line represents the
result for the schematic model with a frequency-dependent second-order
correction, Eq.~(\protect\ref{k2}), included.}
\end{figure}

\begin{figure}
\includegraphics[width=.9\linewidth]{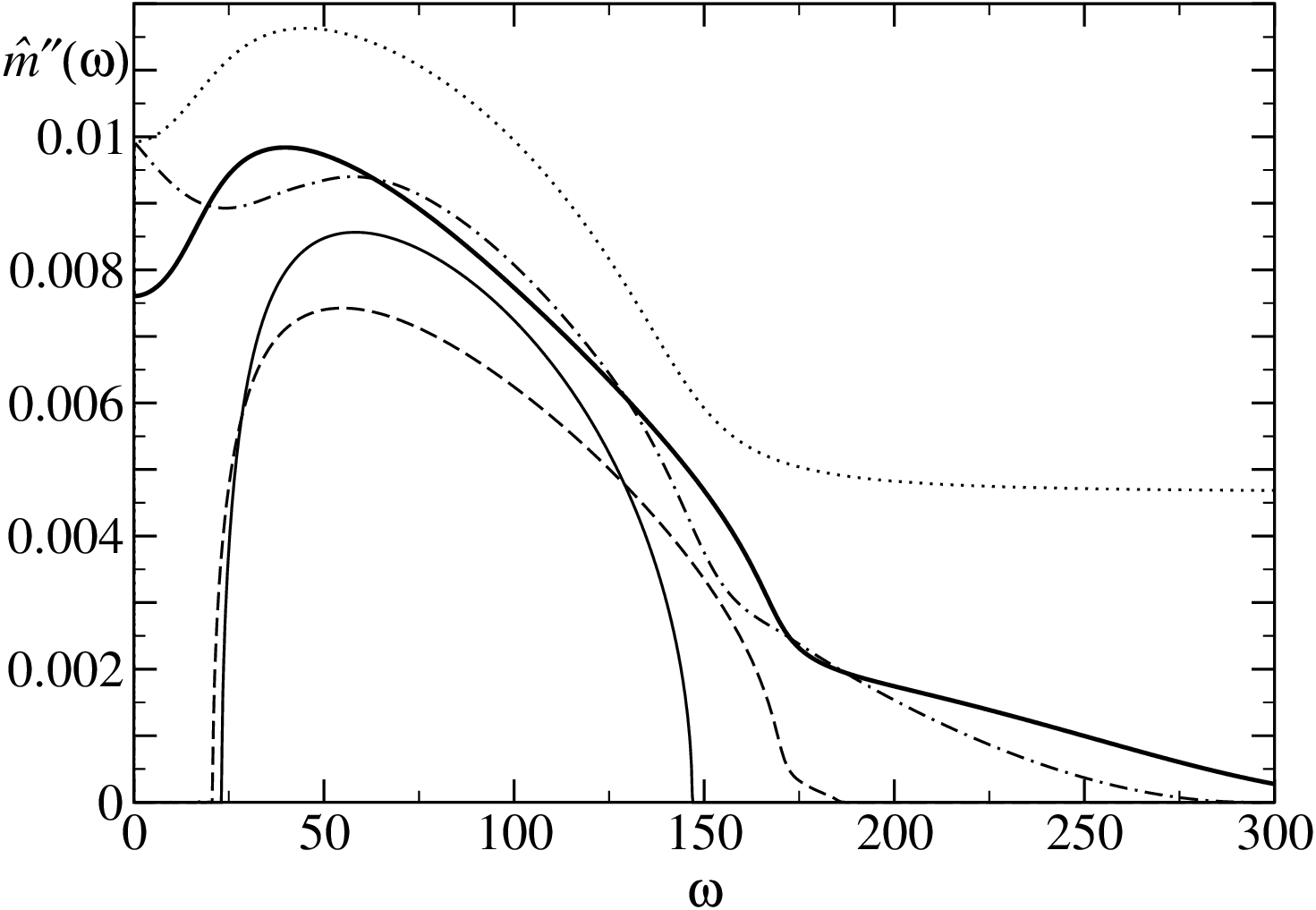}
\caption{\label{fig8}As Fig.~\protect\ref{fig7}, but for $\varphi=0.5676$.}
\end{figure}

\begin{figure}
\includegraphics[width=.9\linewidth]{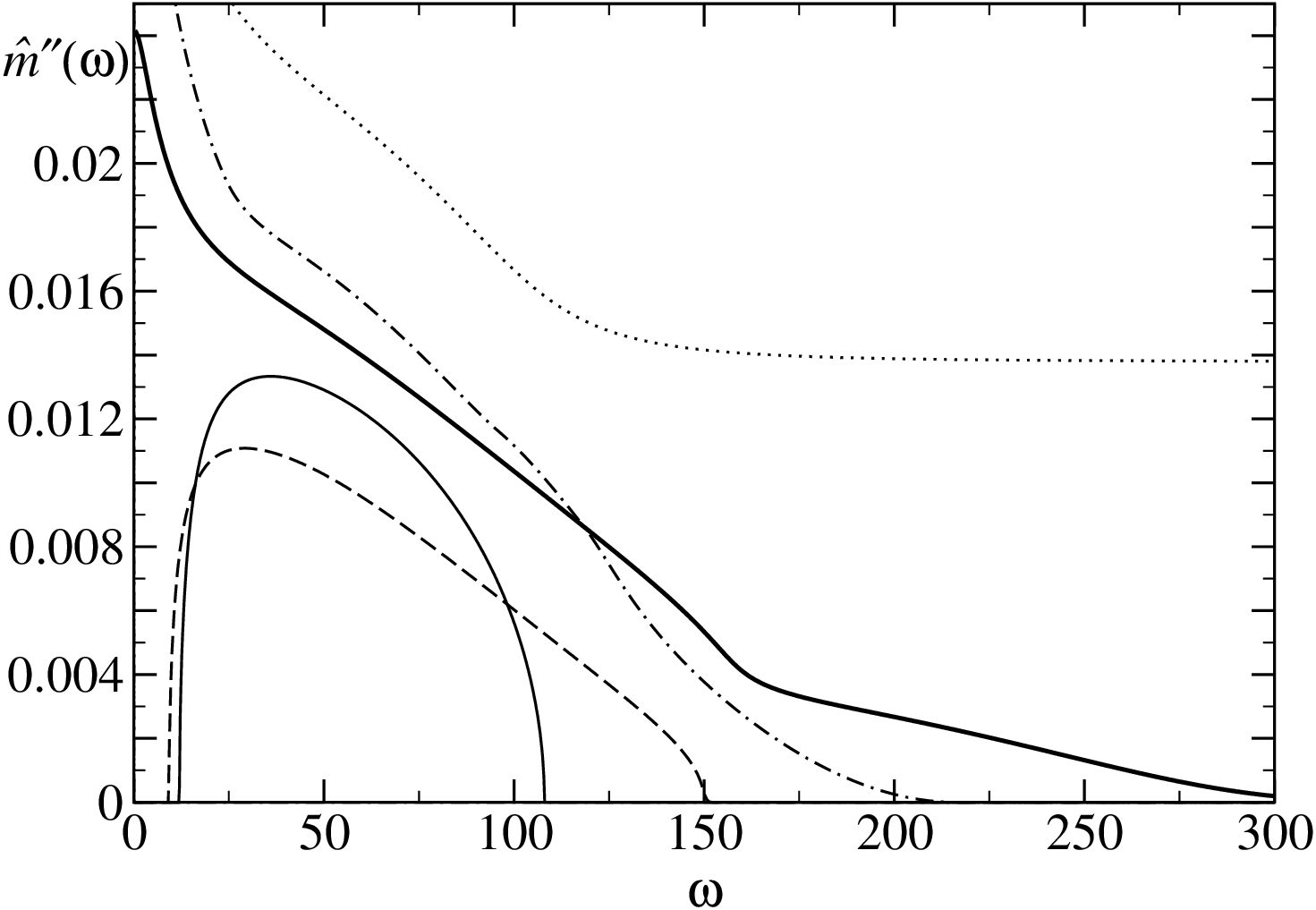}
\caption{\label{fig9}As Fig.~\protect\ref{fig7}, but for $\varphi=0.54$.}
\end{figure}

As a last step, one can also compare the memory kernel spectra resulting from
the SGA in the $q\to0$ limit with the schematic model description proposed
by G\"otze and Mayr, Eqs.~(\ref{schematic}). Fig.~13 of \cite{Goetze2000}
presented a comparison at $\varphi=0.6$ between the
schematic model semi-ellipse and the full-MCT kernel spectrum at
$q=0$. Let us extend the discussion by adding the corresponding SGA spectrum.
This is done in
Fig.~\ref{fig7}, while Figs.~\ref{fig8} and \ref{fig9} carry the same
comparison to lower packing fractions, $\varphi=0.5676$ and $\varphi=0.54$,
respectively. The averaged frequency was chosen to be
$\tilde\Omega=120$ for $\varphi=0.6$, in accordance with \cite{Goetze2000},
and $\tilde\Omega=85$ ($60$) for $\varphi=0.5676$ ($0.54$).
One could improve on these values, but this would not give further physical
insight, nor would it significantly alter the results discussed below.

The success of the semi-ellipse description at $\varphi=0.6$
is based largely on the fact, that in the relevant $q$ range, the variations
of $\hat\Omega_q$ are suppressed as $\eta$ becomes small. For lower packing
fractions, the relative variations in $\hat\Omega_q$ are larger. Together with
the increasing two-mode contributions, this leads to
stronger deviations of the full-MCT spectra from the schematic
semi-ellipse form.

As explained above, the two-mode contributions can within the schematic
description be accounted for approximately
by including a frequency-independent damping term, Eq.~(\ref{k2tau}).
While this procedure works well for $\varphi=0.6$,
it does not lead to an satisfactory result
at the two lower packing fractions investigated here, cf.\ Figs.~\ref{fig8}
and \ref{fig9}.
One needs to go beyond the white-noise approximation and include a
frequency-dependent term, Eq.~(\ref{k2}). Then the spectra are
described fairly, but the advantage of analytic solvability is lost.

However, the SGA solutions include no second-order contributions and thus
should be compared to the pure semi-ellipse of the schematic description.
It is remarkable that the SGA on the other hand
catches the nontrivial shape of the BP-like
contribution to the memory kernel spectrum qualitatively correct even at
lower packing fractions, albeit with an overall too low intensity.
The ``tail'' of the spectrum as well as the evolving central
peak is missing in the SGA solution
due to the well-understood occurence of the pseudo-gap.

\section{Conclusions}\label{sec:conclusio}

The stiff-glass approximation (SGA) to the mode-coupling theory of the
glass transition (MCT) was discussed and shown to provide a good description
of the spectra in the boson-peak regime.
It holds valid even for packing fractions lower than $\varphi=0.6$, where
the analytical formula of G\"otze and Mayr \cite{Goetze2000},
Eq.~(\ref{semi-ellipse}), cannot describe the memory kernel any longer.
This failure of Eq.~(\ref{semi-ellipse}) is
due to increasing variations with $q$ in the phonon
dispersion $\hat\Omega_q$, which in the schematic model is neglected.
The shape of the SGA memory kernel shows more structure than a simple
semi-ellipse form. This leads to a substantially correct description
of high-frequency sound and sound damping in this approximation.

Even the combination of the SGA with the generalized hydrodynamic description
(GHD) gives reasonable results for the memory kernels as well as for the
density correlators, for $q$ vectors where the
damping of the sound mode is dominated by contributions from the
BP-like memory kernel. Outside this window, one gets weakly
damped sound waves, an artifact introduced in the intermediate $q$ range
mainly by the GHD, not the SGA. Nevertheless, the GHD-SGA
gives a basically correct description of the BP
spectra. This finding further corroborates the explanation of the HFS
as a result of the BP spectra.

As a side effect, it was demonstrated, that equations of the form of
Eqs.~(\ref{sga}) can be solved relatively easy in the time domain. At first
glance, this seems paradox, since in Eq.~(\ref{sga}), both $q$ and frequency
$z$ only appear as parameters, while in the time domain one has to solve
a convolution integral.

\begin{acknowledgments}
The author acknowledges stimulating and helpful
discussions with T.~Franosch, M.~Fuchs, W.~G\"otze,
and M.~Sperl. Financial support was partly provided through
DFG grant No.~Go.154/12-1.
\end{acknowledgments}

\bibliography{mct,add}
\bibliographystyle{apsrev}

\end{document}